\documentstyle[12pt,epsf]{article}
  \newlength{\abstractwidth}
\newcommand{\onefigure}[2]{\begin{figure}[htb]
\begin{center}\leavevmode\epsfbox{#1.eps}\end{center}\caption{#2\label{#1}}

 \end{figure}}

  \renewcommand{\thefootnote}{\fnsymbol{footnote}}
  \renewcommand{\thanks}[1]{\footnote{#1}} 
  \newcommand{\starttext}{
  \setcounter{footnote}{0}
  \renewcommand{\thefootnote}{\arabic{footnote}}}
  \renewcommand{\theequation}{\thesection.\arabic{equation}}
  \newcommand{\be}{\begin{equation}}
  \newcommand{\bea}{\begin{eqnarray}}
  \newcommand{\eea}{\end{eqnarray}}
  \newcommand{\beq}{\begin{equation}}
  \newcommand{\ee}{\end{equation}}
  \newcommand{\eeq}{\end{equation}}

  \def\ba{\begin{eqnarray}}
  \def\ea{\end{eqnarray}}

  \def\14{{1\over4}}
  \def\12{{1 \over 2}}
  \def\eq{&=&}

  \def\ds{\partial_{\sigma}}
  \def\h3{h^{3\over 2}}
  
  \def\ds{de Sitter space}

  \def\dsd{de Sitter dual}
  \def\tf{thermofield}
  \def\tfd{thermofield double}
  \def\lb{\label}
  
  \def\bdg{b^{\dag}}
  \def\adg{a^{\dag}}

\begin{document}
  \renewcommand{\theequation}{\thesection.\arabic{equation}}
  \begin{titlepage}
  \bigskip
  \rightline{SU-ITP 01/56}
  \rightline{hep-th/0112204}

  \bigskip\bigskip\bigskip\bigskip

  \centerline{\Large \bf {The Trouble with   }} \centerline{\Large
  \bf {De Sitter Space }}

  \bigskip\bigskip
  \bigskip\bigskip
   \begin{center}
  {\large Naureen Goheer }\\ Department of Mathematics and Applied Mathematics,
 University of Cape Town,\\7701 Rondebosch, Cape Town, South Africa\\
  \end{center}
   \begin{center}
  {\large Matthew Kleban }\\ Department of Physics,
Stanford University\\ Stanford, CA 94305-4060
  \end{center}
\begin{center}
  {\large Leonard Susskind}\\ Korea Institute for Advanced Study and Department of Physics,
  Stanford University\\ Stanford, CA 94305-4060 \\ \vspace{2cm}
  \end{center}

  \bigskip\bigskip
  \begin{abstract}
  In this paper we assume the de Sitter space version of black hole
  Complementarity  which states that a single causal patch of \ds
  \ is described  as an isolated finite temperature
  cavity bounded by a horizon which allows no loss of information.
  We discuss the how  the symmetries of \ds \ should be
  implemented. Then we prove a no go theorem for implementing the
  symmetries if the entropy is finite. Thus we must either give up
  the finiteness of \ds \ entropy or the exact symmetry of the
  classical space. Each has interesting implications for the very
  long time behavior.  We argue that the lifetime of a de Sitter 
  phase can not exceed the Poincare recurrence time.  This is 
  supported by recent results of Kachru, Kallosh, Linde and Trivedi.  

  \medskip
  \noindent
  \end{abstract}

  \end{titlepage}
  \starttext \baselineskip=17.63pt \setcounter{footnote}{0}



\setcounter{equation}{0}
\section{Thermofield Dynamics and Horizons}

An exact formulation of the quantum theory of an accelerating
universe appears to be very elusive \cite{quint}. Thus far the
holographic principle \cite{'thooft,world,raph} has not produced a
dual description of \ds \ analogous to the holographic duality
between anti--de Sitter space and Super Yang Mills theory
\cite{adscft,witten,suswit}

 A number of authors \cite{tw1,banksb,fisch,tw2,boussoN, lisa,birthday,lisamatt,eric}
have argued for a de Sitter complementarity principle similar to
the black hole version \cite{stretch, thooftcom}. While differing in the
details, all versions agree that physics in a single causal patch
of \ds \ should be described as an isolated quantum system at at
finite temperature, and that the thermal entropy of the system
should be finite. From previous experience, especially with Matrix
theory and the AdS/CFT duality, we can expect that the holographic
dual of \ds \ will not look much like classical relativity. Most
likely its degrees of freedom will be very nonlocally connected to
to the local semiclassical description of the space. That raises
an interesting question:  Suppose we were presented with the \ds \
holographic dual. How would we recognize it as such? The answer
for Matrix theory and AdS/CFT was initially through the symmetries,
which exactly matched. Supersymmetry, Galilean symmetry, and
conformal invariance were especially important. Thus we might try
to recognize the \dsd \ by finding that its symmetries include the
$SO(d,1)$ symmetry of $d$-dimensional \ds. However, most of the group (and the most
interesting part of it) involve transformations which take the
causal patch to another patch. The manifest symmetries of the causal
patch are the $SO(d-1)$ rotations which preserve the horizon, and the time
translations. There are another $d-1$ compact generators and $d-1$
noncompact generators which displace the observer to a new patch.
These generators do not act on the Hilbert space of a single
observer. To understand how they act  we have to introduce a
bigger space called the \tfd.

Thermofield theory was invented  \cite{tfd} in the context of many
body theory for the purpose of simplifying the calculation of real
time correlation functions in finite temperature systems. Its
connection with black holes was realized by Israel \cite{israel}
and elaborated in the holographic context by Maldacena
\cite{juan}. Begin with a thermal system characterized by a
Hilbert space $\cal{H}$, a hamiltonian $H$ and a temperature
$T=1/{\beta}$. The thermal ensemble is described by the Bolzmann
density matrix
\be
\rho = {1\over Z}\exp{(-\beta H)}.
\label{rho}
\ee
Thermal expectation values of real time correlation functions
(response functions) contain information not only about
equilibrium properties but also  non--equilibrium dynamics. A
typical response function can be thought of as determining the
future behavior of a system that has been kicked out of
equilibrium. A typical example has the form
\bea
\langle A(0) B^\dag(t)\rangle \eq {1\over Z}Tr e^{-\beta H} A e^{iHt}B
e^{-iHt}\cr
\eq \sum_{ij}A_{ij}B^\dag_{ji}e^{i(E_i-E_j)t-\beta E_i}.
\label{correlator}
\eea
In fact the full set of thermal real time correlators
contains information about   physics arbitrarily far from
equilibrium.

The thermofield formalism introduces a fictitious system that
includes two copies of the original system. The two copies are
labeled $1$ and $2$. Copy $1$ is thought of as the real system,
while $2$ is introduced as a trick. The doubled system is
called the \tf \ double. The Hilbert space for the \tf \ double is
${\cal{H}}_{tf}={\cal{H}}_1\bigotimes {\cal{H}}_2$ where each
factor space is identical to the original Hilbert space. The
Hamiltonian for the \tfd \ is
\be
H\equiv H_{tf}=H_1-H_2.
\label{Hdiff}
\ee
Notice that the energy eigenvalues for  subsystem $2$ are of
opposite sign from those of $1$. We now construct the entangled
state
\be
|\psi\rangle= {1 \over Z^{\12}}  \sum_i e^{-\12\beta E_i}|E_i,E_i\rangle,
\label{psi}
\ee
where the state $|E_i,E_j\rangle$ means the eigenvector of $H_1$
and $H_2$ with eigenvalues   $E_i,E_j$. The state $|\psi\rangle$
is a particular eigenvector of $H_{tf}$ with eigenvalue zero.
Obviously $|\psi\rangle$ has been constructed so that the density
matrix for subsystem $1$ is just the thermal density matrix
$\rho$ at temperature $\beta$.

Operators which belong to subsystem $1$
have the form $A_1\bigotimes I_2$, and will be called $A_1$.
Operators associated with subsystem $2$ will be defined
in a similar manner except with an additional rule of hermitian conjugation:
\be
A_2\equiv I_1\bigotimes A^{\dag}_2.
\label{a2}
\ee

The  correlation function in \ref{correlator}
may be written as an expectation value:
\be
\langle \psi|A_1(0) B^\dag_1(t) |\psi\rangle.
\lb{tfcor}
\ee

Although no physical significance is ordinarily  attached to
correlators involving $A_2$, we may formally define them; for
example
\be
\langle \psi|A_1(0) B_2(t) |\psi\rangle
\lb{cor12}.
\ee
Note that in both \ref{tfcor} and \ref{cor12}
the time dependence of the operators can be defined via the
Heisenberg picture, using the full \tf \ Hamiltonian $H_{tf}$. Let
us examine  expression \ref{cor12} more carefully. It can be written
explicitly in the form
\be
\langle \psi|A_1(0) B_2(t) |\psi\rangle= \sum_{ij}e^{-({\beta\over2}E_i+{\beta\over2}E_j)}
e^{i(E_i-E_j)t}A_{ij}B^\dag_{ji}.
\lb{tfc1}
\ee
This is similar, but not identical, to \ref{correlator}. The
similarity between the two can be made more apparent by writing
\ref{tfc1} as
\bea
\langle \psi|A_1(0) B_2(t) |\psi\rangle&=& \sum_{ij}e^{-{\beta\over2}E_i}
e^{i(E_i-E_j)(t-i{\beta\over2})}A_{ij}B^\dag_{ji}  \cr
\eq \langle A(0)B(t-i{\beta\over2})\rangle.
\lb{tfc}
\eea
This demonstrates that correlators involving both copies
can be expressed as analytic continuations of ordinary thermal
correlators.

We will now consider the relationship between thermofield dynamics
and quantum field theory in spaces with horizons. The simplest
example involves Rindler space. One plus one dimensional  Minkowski space
can be divided into four quadrants; I, II, III and IV (see Figure 1). Quadrants I and III
consist of points separated from the origin by space--like
separation, while II and IV are displaced in by timelike intervals.
Quadrant I is Rindler space, and can be described by
means of the metric
\be
ds^2=r^2 dt^2-dr^2,
\label{rind}
\ee
where $r$ is proper distance from the origin, and $t$ is
dimensionless Rindler time. The Rindler quadrant may be described
by the Unruh thermal state with temperature
\be
T_{rind} ={1\over 2 \pi} ={1\over \beta_{rind}}
\label{trind}.
\ee

\onefigure{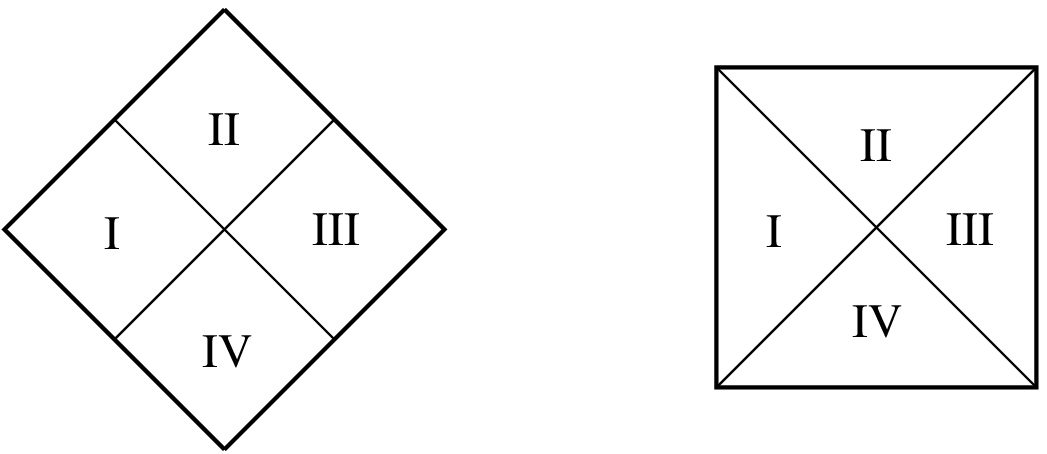}{Penrose diagrams for Minkowski space (on the left) and de Sitter space or the
eternal anti-de
Sitter black hole (on the right).}

Quadrant III is a copy of quadrant I, and can be precisely
identified with the \tfd . To see this we first of all note that
the Rindler Hamiltonian is the boost generator. Since the
Minkowski vacuum is boost invariant, it is an eigenvector of the boost
generator with vanishing eigenvalue. Furthermore, the Minkowski vacuum
is an entangled state of the degrees of freedom in the two
quadrants I and III. Finally, it is well known that when the
density matrix for quadrant I is obtained by tracing over III
the result is a thermal state at the Rindler temperature.
It is easy to see that correlators between fields in
quadrants I and III are related by exactly the same analytic
continuations derived from \tf \ dynamics. To see this, recall that
the usual Minkowski variables $X^0,X^1$ are related to the Rindler
coordinates by
\bea
X^0 \eq r \sinh{ t} \cr
X^1 \eq r \cosh{t}.
\label{X}
\eea
Since the inverse  Rindler temperature \ref{trind} is $2\pi$, the
continuation in equation \ref{tfc} is
\be
t \to t-i\pi,
\label{cont}
\ee
or, from \ref{X}, $X^{\mu} \to -X^{\mu}$. Thus the \tf \
continuation takes quadrant I to quadrant III.

The symmetry of Rindler space includes only the Rindler time
translations. The choice of origin implicit in the identification
of  Rindler space breaks the symmetries of Minkowski translations.
However, once it is realized that the Minkowski vacuum coincides
with the \tfd\ $|\psi\rangle$, it becomes clear that the action of the
translations can be represented in the product Hilbert space ${\cal{H}}_{tf} =
{\cal{H}}_1\otimes{\cal{H}}_2$. The full Poincare algebra in $1+1$ dimensions is
\bea
[P_0,H_{rind}]\eq iP_1 \cr
[P_1,H_{rind}]\eq iP_0 \cr
[P_0,P_1]\eq0.
\label{rindgroup}
\eea

There is one  discrete symmetry which acts on the \tfd ; namely,
the parity operation $X^1 \to -X^1$. This transformation as well
as the transformations generated by $P^{\mu}$ obviously
mix the degrees of freedom of the \tfd \ in a non trivial way.
The lesson of this section is that the full symmetry of
geometries with horizons acts on the \tfd \ Hilbert space, and not
on the space of available to an observer one one side of the
horizon. Moreover the symmetry transformations mix the degrees of
freedom of the two copies.

With this in mind let us consider the eternal AdS eternal black
hole. Maldacena \cite{juan} has noted that the eternal black hole
geometry has two boundaries, one in quadrant I of the Penrose
diagram and one in quadrant III. Maldacena argues that the eternal
black hole should be described by two copies of the usual boundary
conformal field theory, and that the two copies are nothing but the
\tfd \  \footnote{An extremely interesting program is being
pursued by Kraus, Ooguri and Shenker in which  thermofield
correlators are used to probe physics behind black hole horizons
in quadrants II and IV.}. Moreover the state $|\psi\rangle$ is the
Hartle Hawking state.

 Typically there are symmetries which do not mix the two copies
of the \tfd \ and symmetries which do mix them. In the case of the
eternal AdS black hole the symmetry which preserves the separate
boundaries includes time translations and rotations of space. The
thermal  state describing a single copy is not an eigenvector of
either the energy or the angular momentum operators of copy $1$.
However the Hartle Hawking \tfd \ state  $\psi$ is an eigenvector
with vanishing eigenvalue of $H_{tf}$ and $J_{tf}\equiv J_1+J_2$.
As in the Rindler case there is also a discrete symmetry which
interchanges the copies. The Hartle Hawking state is invariant
under this transformation as well.

\setcounter{equation}{0}
\section{de Sitter Space}

de Sitter space is another example of a space with a horizon. Its
Penrose diagram is identical to that of the AdS black hole.
For simplicity we will
consider the case of $2+1$ dimensional de Sitter space. The space
can be globally described by the metric
\be
ds^2 = R^2\left(dt^2 -(\cosh{t})^2 d\Omega^2_2
\right).
\label{ds}
\ee

Quadrant I is now referred to as a causal patch of de Sitter
space. Its metric is given by
\be
ds^2/R^2=(1-r^2)dt^2 -(1-r^2)^{-1}dr^2 -r^2 d\theta^2
\label{patch}
\ee

 As  in the black hole case, the properties of the causal
patch are described by a thermal density matrix \cite{gibhawk}.
Unlike the AdS case we do not know the details of the holographic
dual that describes the causal patch, but we will assume that such
a description exists. The assumption that every causal patch has
its own quantum mechanical description as an isolated system is
the analog of Black Hole Complementarity (which
Parikh, Savonije and  Verlinde  refer to  as
Observer Complementarity \cite{eric}). Once again
we can introduce the \tf \ formalism and introduce a copy of the
causal patch representing quadrant III of the de Sitter Penrose
diagram.

The symmetry group of $d$ dimensional \ds \ is the group $SO(d,1)$, which is simply
the Lorentz group of the $d+1$ dimensional embedding space of the de Sitter hyperboloid.
In the case of $d=3$,
there are six generators: three boosts and three rotations.
One of the rotations and one of the boosts preserves the causal
patch; we will refer to these as $J$ and $H$ respectively.
$J$ generates spatial rotations $(\theta \to \theta + const)$,
and $H$ generates time translations. More
precisely $H$ shifts time forward in quadrant I and backward in
quadrant II.  The remaining two boosts $K_1, K_2$ and rotations $R_1, R_2$
do not preserve the causal patch - that is, they mix the two copies
of the \tfd\ in a way analogous to the action of Minkowski translations
on the Rindler wedges.  The generators satisfy the algebra
\bea
[R, J_i] \eq i \epsilon_{i j} J_j \cr
[J_i, J_j]  \eq i \epsilon_{i j} R \cr
[H, J_i] \eq i \epsilon_{i j} K_j \cr
[H, K_i] \eq -i\epsilon_{i j} J_j \cr
[K_i, K_j] \eq - i \epsilon_{ij} R \cr
[J_i, K_j] \eq i \epsilon_{ij} H \cr
[H, R] \eq 0
\label{dsalg}.
\eea
The \tf \ formalism requires $H=H_{tf}=H_1-H_2$, where $H_1 $ acts
on ${\cal{H}}_1$ and $H_2 $ acts on ${\cal{H}}_2$ independently.
The generators $J_i$ rotate the static patch to a new patch; a
 rotation by $\pi$ interchanges the \tfd\ copies. Since the
$J_i$ mix the two copies, it
is evident that they can not be expressed
as a sum or difference of operators in ${\cal{H}}_1$ and
${\cal{H}}_2$. The generators $K_i$ are Hamiltonians for static
patches which are rotated by a relative angle of $\pi/2$. They too mix the degrees of
freedom in the original patch.

The de Sitter group is part of the group of coordinate
transformations, which in general relativity is the gauge group.
For this reason physical states should be invariant under the
action of any of its generators. Note that this refers to
the generators in the \tfd \ and not to the individual copies.
 According to the definition
\ref{psi}, the generator $H_{tf}=H_1-H_2$ annihilates $|\psi\rangle$.
\be
H_{tf}| \psi\rangle=(H_1-H_2)\sum_i e^{-\12\beta
E_i}|E_i,E_i\rangle =0.
 \label{he}
\ee
 A
de Sitter invariant state must also satisfy
\be
J|\psi \rangle=0.
 \label{je}
\ee This is a highly nontrivial
condition but it is part of the definition of  a quantum version
of \ds . Solutions are known to exist. The Hartle Hawking state
for any quantum field theory in a de Sitter background satisfies
\ref{he} and \ref{je}. The condition that $K$ annihilates $\psi$
is of course a consequence of the commutation relations.

\setcounter{equation}{0}
\section{Finite Entropy}

There is one more condition that we must satisfy in order to have a quantum
dual of \ds; the thermal entropy of one
causal patch is finite. The value of the entropy is connected to the size
of the \ds \ by the Gibbons Hawking \cite{gibhawk} formula for de
Sitter entropy:
\be
S={Horizon\ \ Area \over 4G}.
\label{hh}
\ee
For our purposes we can regard \ref{hh} as defining the size of
the \ds \ through its entropy, which we can compute in terms of the
density matrix \ref{rho}:
\be
S=-Tr (\rho \log{\rho}).
\label{rlnr}
\ee
The only thing we will assume is that the entropy is finite. The
consequences of that for the spectrum of $H_1$ are very clear:
the eigenvalues $E_i$ must be discrete.  More precisely, the
number of eigenvalues with energy less than any specified value
must be finite. This condition implies that the spectrum of $(H_1-H_2)$
must be countable.

In the literature much stronger conditions have been assumed for
the spectrum of $H_1$. Banks and Fischler have conjectured that
the Hilbert space of states ${\cal{H}}_1$ should be finite
dimensional. This may be so, but it does not follow from the
finiteness of the entropy. The Entropy is only equal to the
dimensionality of the space of states when the temperature is
infinite. Entropy can certainly be finite even though the Hilbert
space of states is infinite dimensional. We are assuming only the
weaker condition that the spectrum is discrete.

We are now ready to prove a no--go theorem. The finiteness of the
entropy is incompatible with the existence of the symmetry
generators $(H,J_i,K_j)$, and the requirement that
$H$ be a hermetian operator. First, define the hermitian operator
$L \equiv J_1 + K_2 $. It follows from
\ref{dsalg} that
\be
[H,L]=iL.
\label{hl}
\ee
In general, $L$ is not a good operator on the spectrum of $H$.
If it were, \ref{hl} would imply that
$L$ acts as a raising operator on the spectrum of $H$
but would raise the energy by $i$, which of course is inconsistent with the hermeticity of $H$.
However, although the generator $L$ itself is not bounded, $\exp(iL)$
is a group element and a good operator.
Using \ref{hl}, it is easy to see that
\be
e^{iL}(t) \equiv e^{iHt} e^{iL} e^{-iHt} = e^{e^{iHt} iL e^{-iHt}}
=e^{iLe^{-t}}.
\label{Lt}
\ee

We will now assume that the spectrum of $H$ is countable, and use 
the assumption to derive a contradiction.
We have
\be
\left| \langle \alpha |e^{iL}|\alpha \rangle \right| = 1 - \delta.
\label{lessthanone}
\ee
Here $| \alpha \rangle$ is some state in the Hilbert space, and
$\delta > 0$ because $e^{iL}$ is unitary and $L$ is non-zero.
We also have
\be
F(t) \equiv \langle \alpha |e^{iL(t)}|\alpha \rangle 
= \langle \alpha |e^{iHt}e^{iL}e^{-iHt}|\alpha \rangle
=\langle \alpha |e^{iLe^{-t}}| \alpha \rangle.
\label{expec}
\ee
From this, $F(t) \rightarrow 1$ as
$t \rightarrow \infty$, and $F(0) = 1 - \delta < 1$.  We will now prove that
$F(t)$ is quasiperiodic (see e.g. the appendix of \cite{lisamatt}).

Any sum of the form 
\be
\sum_{n=1}^{\infty} f_n e^{i \omega_n t}
\ee
is quasiperiodic if 
\be
\sum_{n=1}^{\infty} \left| f_n \right |^2 < \infty.
\label{criterion}
\ee
Therefore, it suffices to show that $F(t)$ can be written as a sum
of this form.
But, expanding the state $| \alpha \rangle$ in the energy basis,
\be
F(t) = \sum_{n,m} f^*_n f_m \langle n | e^{i L} | m \rangle e^{i (\omega_n - \omega_m)t}.
\ee
Consider the sum

\be
\sum_{m,n} f^*_n f_m f^*_m f_n  \langle n | e^{i L} | m \rangle 
\langle m | e^{-i L} | n \rangle = \sum_n \left| f_n \right| ^2 
\sum_m \left| f_m \right|^2 \langle n | e^{i L} | m \rangle 
\langle m | e^{-i L} | n \rangle.
\ee
Considering the inner sum, we have (since $\sum_m  \langle n | e^{i L} | m \rangle 
\langle m | e^{-i L} | n \rangle = 1$, and the terms are real and positive)
\be
\sum_m \left| f_m \right|^2 \langle n | e^{i L} | m \rangle 
\langle m | e^{-i L} | n \rangle \leq 1,
\ee
and therefore
\be
\sum_{m,n} f^*_n f_m f^*_m f_n  \langle n | e^{i L} | m \rangle 
\langle m | e^{-i L} | n \rangle \leq 1.
\ee
This shows that $F(t)$ satisfies the criterion \ref{criterion}, and hence
$F(t)$ is quasiperiodic.  Therefore, since $F(0) < 1$, $F(t)$ can not tend to $1$ as 
$t \rightarrow \infty$, and we have a contradiction

This proves that $H$ can not have a countable spectrum.
and therefore that the $E_i$ cannot be discrete and the
entropy cannot be finite.

Shenker \cite{steve} has suggested an interesting viewpoint; that of
anomalies. It is fairly clear that in ordinary perturbation theory
in \ds \, the symmetry of the space will never break down.
Furthermore, in every order the entanglement entropy of I and III
will be infinite. This suggests that the same nonperturbative
effects which make the spectrum discrete and the entropy finite,
also break the de Sitter symmetry.  A tantalizing hint is that the
size of such a nonperturbative
effect would scale as $\exp(-{\rm area}/4G_N) \sim \exp(-S)$, which is the
characteristic energy gap $\delta E$ for a system with entropy $S$.

The same argument can be applied to Rindler space. Using
\ref{rindgroup} and replacing the operator $L$ by $P_1+P_0$ the
spectrum $H_{rind}$ is also proved to be all real numbers. Thus
the entropy of Rindler space must be infinite.
 Obviously if the number of
non--compact directions of spacetime is 3 or more, the horizon is
infinite and the total entropy does diverge. It is only in the
case of $1+1 $ non--compact directions that the entropy could
possibly be finite and conflict with Rindler symmetry. However, at
least in the case of string theory it is known
 that compactification
 to $1+1$ dimensions can not yield a theory with translation invariance
\cite{banksuss}.  In the appendix we describe as an example the action
of a Minkowski space translation on the Rindler thermofield double.

\setcounter{equation}{0}
\section{Conclusions}

Finiteness of entropy appears to be incompatible with de Sitter
symmetry.
What are we to make of this no--go theorem? One
possibility is that \ds \ has infinite entropy.
Perhaps only
entropy differences are finite.  However, this seems
to run against the grain of everything we have learnt over the
last decade. The horizon of \ds \ is locally identical to that of
a Schwarzschild black hole. It is hard to see why one would have
an infinite entropy and the other not \cite{edi}.

If the entropy is finite then the symmetry of different causal
patches must be broken, the Hamiltonians and energy spectra
differing at least slightly. How would these effects manifest themselves?
The discreteness of
the energy spectrum introduces a new time scale of order
$1/\delta$, where $\delta$ is of order the typical spacing of
levels. For a system with entropy $S$ the level spacing is of
order $\exp{(-S)}$ and the time scale is the Poincare recurrence
\cite{lisa,birthday,lisamatt} time  $t_p \sim \exp{S}$. This is
the time scale on which the discreteness of the spectrum causes
significant effects. Thus the violations of de Sitter symmetry
should become important for times of order  $t_p$ and  recurrences
should not respect the symmetry. We may speculate that no space
can behave like idealized \ds \ for times longer than $t_p$.
For example, it is possible that eternal de Sitter simply does not exist,
because there are always instabilities which cause the space to
decay in a time shorter than the recurrence time.\footnote{Some time ago,
Simeon Hellerman
\cite{simeon} suggested that there may be a general bound
 coming from an acceleration--duration
relation, which would prohibit arbitrarily long-lived accelerating spacetimes
such as de Sitter.}

An alternative has been proposed by Bousso \cite{raph}, which is simply
that because any observer will be destroyed by thermal particles
long before $t_p$, times longer than $t_p$ should be regarded as having no
operational meaning.  In the same spirit,
Banks, Fischler, and Paban \cite{tw2} speculate
that due to the physical constraints on the measuring process in
\ds \ , no experiment can ever last a time comparable to the
recurrence time. Hamiltonians that agree for time scales less than
$\sim t_p$ should be considered physically equivalent. In terms of the
energy levels $E_i$ this would mean that the level shifts going
from one patch to another
should be no bigger than the level spacing $\exp{(-S)}$. If this
can be made precise and the
physical consequences of different Hamiltonians are identical,
then the exact choice of Hamiltonian would become part of the
gauge redundancy in a single patch. The transformation from one
patch to another would include such a gauge transformation.
It is not clear to us whether this would mean that Poincare
recurrences are meaningless or if it means that there is an
additional degree of
unpredictability in their occurence.

\setcounter{equation}{0}
\section{Note Added in Revised Version}

We have argued on general grounds that de Sitter behavior can not
persist for times of order the Poincare recurrence time. In this
note we will present an argument based on work by Kachru, Kallosh,
Linde and Trivedi \cite{linde} that strongly supports this claim  
in the context of string theory \footnote{We are especially
grateful to Andrei Linde for a very enlightening discussion of
this point and for communicating the results of \cite{linde} to   
us. }. Since string theory has supersymmetric ground states with
vanishing vacuum energy it is possible to tunnel out of any
positive  local minimum of the scalar potential. Typically there
will be a barrier to the tunneling which may be as high as the   
string or Planck scale.  Kachru, Kallosh, Linde and Trevedi
compute the rate for  tunelling over such a barrier and find that
it is always less than the recurrence time. The essence of the
calculation is simple and can be understood by the following (over)simplified
argument:

Obviously it is sufficient to calculate the rate for a fluctuation
that takes the static patch to the top of the potential barrier
since once that configuration is achieved, there is no further
obstacle to rolling down to the vacuum with vanishing energy.
Suppose the vacuum energy density (in Planck Units) at the top of
the barrier is $\lambda <1$. If the barrier is broad and fairly  
flat, we will have a temporary de Sitter phase with horizon area
$\sim 1/{\lambda}$ and similar entropy. Now consider the rate for 
a thermal fluctuation to take take the static patch from the local
minimum to the top of the potential. The standard formula for the
rate of thermal fluctuations $\gamma$ is $$ \gamma ~
\exp{(S_1-S_0)}. $$ where $S_0$ and $S_1$ are the entropies of the
equilibrium phase and the fluctuation respectively. Since the   
recurrence time is $\exp{S_0}$ it is clear that the rate of
thermal tunelling is always bigger than the rate of Poincare
recurrences.  Thus is appears that in string theory, a metastable
de Sitter phase can not live for a recurrence time.

This instability has implications for the energy spectrum of the 
metastable de Sitter.  Since the lifetime is always shorter than 
the recurrence time, each energy level must develop a width $\gamma$
at least of order $\exp{-S}$.  In a typical situation of the type
studied in \cite{linde}, the width of each level is vastly greater
than the level spacing, which is reminiscent of a Schwarzschild black 
hole in flat space.

\setcounter{equation}{0}

\section{Acknowledgements}
N. G. would like to thank C.W.Kim and the staff of KIAS for
hospitality while this work was in progress, and UCT for supporting her
 studies at UCT.  M. K. is the Mellam Family Graduate Fellow.

M.K. and L.S. are grateful for discussions with Tom
Banks, Puneet Batra, Raphael Bousso, Nemanja Kaloper, Andy Albrecht, Simeon
Hellerman, Steve Shenker, and especially Ben Freivogel.

\setcounter{equation}{0}
\section{Appendix}

In this appendix we illustrate how the group of Minkowski translations acts 
on the Rindler degrees of freedom. For simplicity we study a 1+1 Minkowski 
toy model with a free scalar field $\zeta$ of mass $m$. The Minkowski 
coordinates are denoted by $T, ~X$ and take values in $[-\infty,\infty]$. 
The Lagrangian density for a free scalar field of mass $m$ is given by
\be
{\cal L}= \sqrt{-g}~[g_{\mu \nu} {\partial}^{\mu}\zeta
{\partial}^{\nu}\zeta -m^2 \zeta],
\ee
Rindler coordinates are defined by
\be
r=\sqrt{X^2-T^2}, ~ \tau=\frac{1}{2} \ln{\frac{X+T}{X-T}}. \ee
They are defined on both the right and the left Rindler wedge, but
time and radius run in opposite directions in the two wedges.
In these coordinates the wave equation takes the form
\be
\ddot{\zeta}-\frac{\partial ^2}{{\partial u}^2} \zeta +m^2e^{2u}
\zeta=0 \label{mfldeq} \ee where $u=\ln{r}$ and dot indicates derivative
with respect to $\tau$. 
Let us note first that as we discussed earlier, the algebra of the Rindler 
Hamiltonian and the Minkowski translations can only be consistent if the 
entropy of Rindler space is infinite. Since the $u$ axis is unbounded at
both ends it is not surprising that the entropy of the scalar
field at finite temperature is infinite. However the important
infinity comes from one end as can be seen from the wave equation 
(\ref{mfldeq}): the mass term acts like a
wall preventing the field from penetrating to large positive $u$.
On the other hand the region of large negative $u$ is still
massless. In this case the divergent entropy comes from the
$u<<0$. This is of course the region very near the horizon.

Any solution of the wave equation can be expanded in terms of
the basis functions
\be
f_k=\frac{1}{\sqrt{c_k}}e^{iku-i{\omega}_k\tau}=\frac{1}{\sqrt{c_k}}r^{ik}
e^{-i\omega_k \tau} \ee where $c_k=4 \pi \omega_k$ are
normalization constants and $\omega_k=+\sqrt{k^2 + r^2 m^2}$.

The two \tf \ copies of $\zeta$ at $\tau =0$ are just the fields on the two
half spaces $X>0$ and $X<0$. Each copy can be expanded in creation
and annihilation operators for Rindler excitations. Denoting the operators 
by $a, a^{\dag}$ and $b, b^{\dag}$ for $X>0$ and $X<0$
respectively $\zeta$ can be expressed as 
\be
\zeta(r,\tau)=\int  \frac{dk}{\sqrt{c_k}} [{r}^{ik}e^{-i\omega_k{\tau}} a(k)+
{r}^{-ik} e^{i\omega_k \tau} a^\dagger(k)]\Theta(X) +\int
 \frac{dk}{\sqrt{c_k}}
[r^{ik}e^{-i\omega_k\tau} b(k)+ r^{-ik}e^{i\omega_k \tau}
b^\dagger(k)]\Theta(-X). \ee The first integral corresponds to the
part of the wave equation in the right Rindler wedge while the
second integral corresponds to the part in the left Rindler wedge.

To illustrate the action of the Minkowski translations on the
Rindler degrees of freedom we consider a translation along the X
axis by distance $a$. The Rindler coordinates transform as
\be
r'=\sqrt{(X-a)^2-T^2}=\sqrt{r^2+a^2 \mp 2ar\cosh{\tau}} \ee and
\be
\tau '=\frac{1}{2}\ln{\frac{X-a+T}{X-a-T}}=\frac{1}{2}
\ln{\frac{\pm r \cosh{\tau} +r \sinh{\tau}-a}{\pm r \cosh{\tau} -r
\sinh{\tau}-a}}. \ee The upper signs apply in the right Rindler
wedge ($X>0$), the lower ones in the left Rindler wedge ($X<0$).

In the right hand wedge, the shifted fields $\zeta,\dot{\zeta}$ at
$\tau=0$ have the form \bea \zeta(r)\eq \int  \frac{dk}{\sqrt{c_k}}
[(r-a)^{ik}
a(k)+(r-a)^{-ik} a^\dagger(k)] \Theta(r-a)\cr &&+\int  \frac{dk}{\sqrt{c_k}}
[(a-r)^{ik} b(k)+
(a-r)^{-ik} b^\dagger(k)]\Theta(a-r) \cr \dot{\zeta}(r) \eq -i \int
 \frac{dk}{\sqrt{c_k}}\omega_k r[(r-a)^{ik-1} a(k)
-(r-a)^{-ik-1}a^\dagger(k)]\Theta(r-a)\cr
 &&-i \int  \frac{dk}{\sqrt{c_k}} \omega_k r[(a-r)^{ik-1} b(k)- (a-r)^{-ik-1}
 b^\dagger(k)]\Theta(a-r).
 \label{shifd}
\eea
The shifted field can also be to expanded in terms of new
(shifted) creation and annihilation operators $c^\dagger$ and $c$.
Concentrating on the right hand Rindler wedge:
\be
\zeta(r,\tau)\mid _{\tau=0}=\int dk [f_k c(k)+f_k^* c^\dagger(k)] \ee
where the basis functions $f_k$ are defined as above.

The Bogoliubov transformation giving the operators $c,c^{\dag}$ is
easily obtained. For example we find
\be
 c^\dagger(l)=\int dk~ [D(l,k) \adg (k) +E(l,k)a(k)+F(l,k)\bdg
 (k)+G(l,k)b(k)]
\ee with the Bogoliubov coefficients given by
\bea D(l,k) \eq
\frac{1}{4\pi\sqrt{\omega_k \omega_l}}(\omega_k+\frac{k}{l}~\omega_l) 
a^{il-ik}B(-il+ik,-ik)  \cr
E(l,k) \eq
-\frac{1}{4\pi\sqrt{\omega_k \omega_l}}(\omega_k+\frac{k}{l}~\omega_l) 
a^{il+ik}B(-il-ik,ik) \cr
F(l,k) \eq
\frac{1}{4\pi\sqrt{\omega_k \omega_l}}\frac{l\omega_k-k\omega_l}{l-k} 
a^{il-ik}B(il,-ik) \cr
G(l,k) \eq
-\frac{1}{4\pi\sqrt{\omega_k \omega_l}}\frac{l\omega_k-k\omega_l}{l+k} 
a^{il+ik}B(il,ik), \eea
where $B(a,b)$ is the Euler $\beta$--function:
\be
B(a,b)=\int_0^1  y^{a-1}(1-y)^{b-1} dy \ee and we have used the identity 
$B(a,b+1)=\frac{b}{a+b}B(a,b)$. 
In the massless limit these coefficients 
simplify to give
\bea D(l,k) \eq
\frac{1}{2\pi}\sqrt{k \over l} a^{il-ik} B(-il+ik,-ik)~ \Theta (kl) \cr
E(l,k) \eq - \frac{1}{2\pi}\sqrt{{k \over l}}
a^{il+ik} B(-il-ik,ik)~ \Theta (kl) \cr
F(l,k) \eq (sgn(l)-sgn(k))\frac{\sqrt{|kl|}}{4\pi (l-k)}
a^{il-ik}B(il,-ik) \cr
G(l,k) \eq -(sgn(l)-sgn(k))\frac{\sqrt{|kl|}
}{4\pi(l+k)} a^{il+ik}
B(il,ik). \eea
Similar equations
determine the transformed operators for the left wedge of Rindler
space. In this way we explicitly see how the symmetries of
Minkowski space mix the two copies of the \tfd .


\begin{thebibliography}{99}

\bibitem{quint} S. Hellerman, N. Kaloper and L. Susskind,
``String Theory and Quintessence,'' JHEP {\bf 0106} (2001) 003,
hep-th/0104180.

 W. Fischler, A. Kashani-Poor, R. McNees and S. Paban,
 ``The Acceleration of the Universe, a Challenge for String Theory,''
JHEP {\bf 0107} (2001) 003, hep-th/0104181.


\bibitem{'thooft} G. 't Hooft, ``Dimensional Reduction in Quantum Gravity,''
gr-qc/9310026.


\bibitem{world} L. Susskind,  ``The World as a Hologram,'' hep-th/9409089.

\bibitem{raph} R. Bousso, ``The Holographic Principle,"
Rev. Mod. Phys. {\bf 74} (2002) 825-874, hep-th/0203101.

\bibitem{adscft} J.~Maldacena, ``The large N limit of
superconformal field theories and supergravity,'' Adv.\ Theor.\
Math.\ Phys.\  {\bf 2}, 231 (1998) Int.\ J.\ Theor.\ Phys.\  {\bf
38}, 1113 (1998), hep-th/9711200.





\bibitem{witten} E.~Witten, ``Anti-de Sitter space and
holography,'' Adv.\ Theor.\ Math.\ Phys.\  {\bf 2}, 253 (1998),
hep-th/9802150.


\bibitem{suswit} L. Susskind and E. Witten,
``The Holographic Bound in Anti-de Sitter Space,"  hep-th/9805114/



\bibitem{stretch} L. Susskind, L. Thorlacius, and J. Uglum,
``The Stretched Horizon and Black Hole Complementarity,'' Phys.
Rev. {\bf D48} (1993) 3743, hep-th/9306069.

\bibitem{thooftcom} G. 't Hooft, ``Quantum Information and
Information Loss in General Relativity," gr-qc/9509050.



\bibitem{tw1} T. Banks and W. Fischler,
``M-theory Observables for Cosmological Space-times,''
hep-th/0102077.

\bibitem{banksb} T. Banks, ``Cosmological Breaking of Supersymmetry
or Little Lambda Goes Back to the Future II," hep-th/0007146

\bibitem{fisch} W. Fischler, ``Taking de Sitter Seriously,"
talk given at {\it Role of Scaling Laws in Physics and Biology
(Celebrating the 60th Birthday of Geoffrey West)}, Santa Fe, Dec.
2000.

\bibitem{tw2}  T. Banks, W. Fischler and S. Paban,
``Recurrent Nightmares?: Measurement Theory in de Sitter Space,"
hep-th/0210160.

\bibitem{boussoN} R. Bousso, ``Positive vacuum energy and the N-bound,"
JHEP 0011 (2000) 038, hep-th/0010252.


\bibitem{lisa}  L. Dyson, J. Lindesay, and L. Susskind,
``Is There Really a de Sitter/CFT Duality,'' hep-th/0202163.

\bibitem{birthday} L. Susskind,
``Twenty Years of Debate with Stephen,'' hep-th/0204027.

\bibitem{lisamatt}  L. Dyson, M. Kleban, and L. Susskind,
``Disturbing Implications of a Cosmological Constant," JHEP 0210
(2002) 011, hep-th/0208013.

\bibitem{eric}  M. Parikh, I. Savonije, and E. Verlinde,
 ``Elliptic de Sitter Space: $dS/Z_2$,"
hep-th/0209120.

\bibitem{gibhawk} G. Gibbons and S. Hawking,
``Cosmological Event Horizons, Thermodynamics, and Particle
Creation,'' Phys.Rev. {\bf D15} (1977) 2738.

\bibitem{juan} J. Maldacena,
``Eternal Black Holes in AdS,'' hep-th/0106112.



\bibitem{israel}W.~Israel, ``Thermo Field Dynamics Of Black Holes,''
Phys.\ Lett.\ A {\bf 57}, 107 (1976).





\bibitem{tfd} P. Martin and J. Schwinger, Phys. Rev. {\bf 115}, 1342
(1959); J. Schwinger, JMP vol 2, 407(1961);
 K.T. Mahanthappa, Phys.Rev.vol 126, 329(1962);
P.M. Bakshi and K.T. Mahanthappa, JMP vol 4, 1 and 12 (1963);
Keldysh  JETP Vol 47, 1515(1964); Y. Takahashi and H. Umezawa,
Collective Phenomena {\bf 2} 55 (1975).

\bibitem{banksuss}T. Banks and L.Susskind,
``The Number of States of Two Dimensional Critical String Theory,"
Phys.Rev. D54 (1996) 1677-1681, hep-th/9511193.

\bibitem{steve} S. Shenker, private communication.

\bibitem{simeon} S. Hellerman, private communication.

\bibitem{edi} E. Halyo, ``De Sitter Entropy and Strings," hep-th/0107169.

\bibitem{linde} S. Kachru, R. Kallosh, A. Linde and S. Trivedi, ``de Sitter
Vacua in String Theory," hep-th/0301240.


\end{thebibliography}
\end{document}